\documentclass[preprint, showpacs,preprintnumbers,amsmath,amssymb,nofootinbib]{revtex4}
\usepackage{amssymb}
\usepackage{natbib}
\usepackage{graphicx}
\usepackage{dcolumn}
\usepackage{bm}

\newcommand{\bea}{\begin{eqnarray}}
\newcommand{\eea}{\end{eqnarray}}

\begin{document}
\title{ Testing Cosmic Distance-Duality Relation from Future Gravitational Wave Standard Sirens}
\author{Xiangyun Fu$^1$\footnote{corresponding author:  xyfu@hnust.edu.cn}, Lu Zhou$^1$ and Jun Chen$^2$ }
\address{ $^1$Institute of  Physics, Hunan University of Science and Technology, Xiangtan, Hunan 411201, China\\
 $^2$School of Science, Kaili University,
Kaili, Guizhou 556011, China}

\begin{abstract}
 A validation of the cosmic distance-duality relation (CDDR) is crucial because any observational  departure from it could be  a signal of new physics. In this work, we explore the potentialities of luminosity distance data from the gravitational wave (GW) standard sirens of future Einstein Telescope (ET) to test the CDDR.  The angular diameter distance data are used from the galaxy clusters samples and the baryon acoustic oscillation (BAO) measurements. The basic advantage of  GW measurements substituting for the observations from the type Ia supernovae (SNIa) is that the luminosity distance from it is insensitive to the non-conservation of the number of photons. By simulating 550 and 1000 data points of future GW measurements  in the low redshift range $0<z<1$, we show that the measurements of future GW events will be  a powerful tool to test the CDDR.

 $\mathbf{Keywords:}$  Cosmic distance-duality relation,  gravitation wave, galaxy cluster
\end{abstract}


 \maketitle

\section{Introduction}
The cosmic distance-duality relation (CDDR), which relates luminosity distance (LD) $D_{\rm L}$  and
angular diameter distance (ADD) $D_{\rm A}$ to a given source at redshift $z$ through the following identity
\begin{equation}\label{ddr}
  \frac{D_{\rm L}}{D_{\rm A}}{(1+z)}^{-2}=1,
\end{equation}
was firstly proved by Etherington in 1933~\cite{eth1933} based on two fundamental hypotheses, namely, that light travels always along null geodesics in a
Riemannian geometry, and the number of photons  is conserved~\cite{ellis1971,ellis2007}.
 This equation  has been used, without any doubt, in astronomical observations and modern cosmology, since it is independent of Einstein field equations and the nature of matter. However, the possibilities of the CDDR violation may be resulted from non-conservation of the number of photons or a non-metric theory  of gravity, in which the light does not travel along null geodesics~\cite{uzan2004,Santana2017}.  The non-conservation of the total number of photon may be resulted by the presence of scattering and absorption of some opacity sources~\cite{Csaki2002,Lima2011} or non-standard mechanisms such as scalar fields with a non-minimal coupling to the electromagnetic Lagrangian~\cite{Hees2014,Holanda2017,Holanda20161,Aguirre1999} or oscillation of photons propagating in extragalactic magnetic fields into light axion~\cite{Csaki2002,Avgoustidis2009,Avgoustidis2010,Jaeckel2010}. Any violation of CDDR from astronomical observations may even be considered as a signal of exotic physics or the existence of some unaccounted errors in the observations~\cite{bassett}. Thus,  testing  the validity of this relation with different  observational data sets and methods is worthy and necessary.

To check the validity of the CDDR with astronomical observations, one should in principle obtain the LD and ADD of some objects at the same redshift. The LD can be generally obtained through the observation of the type Ia supernovae (SNIa) standard candles, and the ADD can be estimated from the observation of galaxy clusters, the cosmic microwave background (CMB) and baryon acoustic oscillation (BAO). Since the redshifts of LD and ADD data points in the present observations usually do not match, some tests on the CDDR are performed through comparing the observed values with the corresponding theoretical ones from an assumed cosmological model. With the LDs from $\Lambda$CDM model, Uzan {\it et al.}~\cite{uzan2004} and De Bernardis {\it et al.}~\cite{DeBernardis2006}   tested successively the CDDR with galaxy cluster samples~\cite{Reese2002,Boname06}, and found no violation from the CDDR. Then, combining the SNIa data with the standard rulers from the CMB and  BAO measurements, Lazkoz {\it et al.} verified the validity of the CDDR at the $2\sigma$ confidence level (CL)~\cite{Lazkoz2008}. Using the galaxy cluster data from the elliptical and spherical $\beta$ models~\cite{Boname06,DeFilippis05}, Holanda {\it et al.} obtained that the CDDR is compatible with the elliptical and spherical $\beta$ models at $1\sigma$ and $3\sigma$ CL, respectively~\cite{holanda2010}.

Recently, in order to match the redshifts of the galaxy cluster samples~\cite{Boname06,DeFilippis05} with those of  SNIa data by the cosmological-model-independent method, Holanda {\it et al.}~\cite{holanda20103} adopted a criterion ($\Delta z=|z_{\rm ADD}-z_{\rm SNIa}|<0.005$) and chose the closest one. From the Constitution SNIa compilation~\cite{Hicken2009}, they found that the CDDR is marginally consistent with the elliptical $\beta$ model at $2\sigma$, but, it indicates a strong violation from the spherical $\beta$ model even at $3\sigma$ CL. Using the Union 2 SNIa compilation, Li {\it et al.} also performed tests on the CDDR, and found that the CDDR is consistent with the elliptical $\beta$ model at $1\sigma$ CL, and the spherical $\beta$ model at $3\sigma$~\cite{Li2011}. In order to avoid larger statistical errors brought by using merely one SNIa data point from all those available which meet the selection criterion, Meng {\it et al.}~\cite{Meng2012}, instead of using the nearest point of  SNIa compilation,
 binned these data available to obtain LD to match the corresponding ADD sample. They studied  different morphological models of galaxy clusters, and found that the marked triaxial ellipsoidal model is a better geometrical hypothesis to describe the structure of galaxy cluster than the spherical model if the CDDR is valid.
Then, Wu {\it et al.} tested the CDDR by combining the Union 2.1 compilation  and five ADD data points from the BAO measurements, and found that the BAO measurement is a very powerful tool to test the CDDR due to the precision of the BAO measurements~\cite{Wu2015}. Still some other tests, involving the ADD of galaxy clusters~\cite{Boname06,DeFilippis05}, current cosmic microwave background (CMB) observations~\cite{Ellis2013},  Hubble parameter data $H(z)$ from cosmic chronometers, gas mass fraction measurements in  galaxy clusters~\cite{Goncalves2012},  strong gravitational lensing (SGL)~\cite{Cao2012,Cao2015} and time delay lenses~\cite{Balmes}, are performed to investigate the validity of the CDDR by assuming a deformed  CDDR, such as ${D_{\rm L}}{(1+z)}^{-2}/{D_{\rm A}}=\eta{(z)}$, in different redshifts ranges, and the results show that the CDDR is consistent with the observations at certain CLs~\cite{avtidisgous,Holanda2012a,Santos2015,
Stern2010,Holanda2012,Liao2011,Holanda20171,Liao2016,fuxiangyun,Fu2017,Liang2013,Rana,Ruan}.

It is worthy  noting that the LD from SNIa measurements is dependent on the conservation of the number of photons. Any kind of  violation from the  conservation of the number of photons, such as absorption and  scattering of photon or axion-photon mixing, sensibly imprints its impact on the test of the CDDR~\cite{Tolman1930}.   So a common limitation of the aforementioned tests involving LD from SNIa compilations is that, if the evidence of the CDDR violation $\eta\neq 1$ is obtained, the fundamental reason for the departure may not be known because  the results from these tests are sensitive to the both fundamental hypotheses for the  CDDR.

More recently, the joint detections of the gravitational-wave (GW) event GW170817 with electromagnetic (EM) counterpart (GRB 170817A)
from the merger of binary neutron stars (NSs)~\cite{Abbott,Abbott2,Daz,Cowperthwaite} have opened the new era of multi-messenger cosmology, and it makes for the first time that a cosmic event can be investigated in both EM waves and GWs.
The application of GW information in cosmology was first proposed by Schuts~\cite{Schutz}, who suggested that the Hubble constant can be determined from GW observations using the fact that the waveform signals of GWs from inspiraling and merging compact binaries encode distance information. So, GW sources can be considered as standard sirens in astronomy, analogous to supernovae standard candles.   Unlike the distance estimation from  SNIa observations, one can, from the GW observations, obtain the luminosity distances directly without the need of cosmic distance lader since stand sirens are self-calibrating. This advantage of GW measurements can help us dodge the influence of the non-conservation of the number of photon on the test of CDDR. If compact binaries are NS-NS or black hole (BH)-NS binaries, the source redshift may be observed from EM counterparts that occur coincidentally with the GW events~\cite{Zhao2011,Nissanke2010,Cai2017}. Thus, the LD-redshift relation can be constructed by a cosmological-model-independent way through combining the measurements of the sources' redshifts from the EM counterpart, and it provide us with an opportunity to make constraints on the cosmological parameters and  the possible departures from the CDDR. It is worth mentioning that the propagation of GWs in modified gravity theories is in general different from that in General Relativity and the LDs from GWs are different from that for the electromagnetic
signals~\cite{Belgacem2018,Saltas2014}. Therefore, if one tests the CDDR with LDs from GW measurements and distances from electromagnetic observations, the violation of CDDR might indicate deviations of gravity theory from General Relativity besides the existence of new phsics.  In
this work, the main motivation is to employ the GW  as alternative for the SNIa
 to test CDDR in the frame of General Relativity.

Up to now, the simulated GW data have been used to measure the cosmological parameters~\cite{Zhao2011,Pozzo201217,Liao2017,Cai2016,Wei2017}, determine the total mass of neutrino~\cite{Wang2018}, investigate the anisotropy of the universe~\cite{Cai2018,Wei2018} and make constraints on the evolving Newton's constant G~\cite{Zhao2018}. More recently, Yang {\it et al} explored the potentialities of future GW detections  to constrain a possible departure from the CDDR through combining LD of simulated gravitational wave data from the  Einstein Telescope (ET) and ADD from SGL systems in a relative high redshifts range $z\sim3.6$~\cite{Yang2017}. They obtained that future results from GW data will be at least competitive with current constraints on CDDR from SNIa+GRBs+SGLs analyses. However, it is shown that mass profile of lensed galaxies and dynamical structure may bring forth significant changes in lensing studies~\cite{Cao2012,Cao2015}, and its impact on the test of CDDR needs further investigations.
Thus, one needs more relevant ADD data to explore the potentialities of GW measurements on the test of the validity of CDDR. It is well known that BAO measurement is a very precise astronomical observation~\cite{Bassett2010,Wu2015} and can be used as a very powerful tool to test the CDDR~\cite{Wu2015}. In addition, the measurement of  galaxy clusters also plays an important role to test this relation.  So, it is worth to confirm the ability of   future GW measurements jointly with the ADD data from BAO and galaxy clusters  to constrain a possible departure from this reciprocal relation.

In this work, we detect the potentialities of  future GW measurements to test CDDR. The analyses are carried out with the LD ($D_{\rm L}$) from simulated GW data jointly with ADD from BAO and galaxy cluster samples. We simulate 550 and 1000 data points of GWs from the ET in the redshifts range $0<z<1$ as reference, and  impose limits on   $\eta(z)$ to estimate the possible departures from the CDDR.  In order to compare our results with that from Refs.~\cite{Meng2012,Wu2015},  we also employ the binning method to obtain the corresponding LDs from simulated GW data for each  BAO or galaxy cluster systems.  The results indicate  that measurement of  future GW events will be a very powerful tool to realize the validation of this reciprocal relation.

\section{samples and simulated GW data}
The structure of galaxy clusters is essential for the cosmological probe~\cite{Suwa2003,Allen2004}.  Generally speaking, to get reasonable results of ADD from galaxy cluster observation, one has to assume certain cluster morphologies and employ the joint analysis of SZE and X-ray brightness measurements~\cite{SZE}. Two galaxy cluster samples are utilized to obtain the ADD through different morphological assumptions. The first one involves 25 X-ray-selected galaxy clusters~\cite{DeFilippis05} described as an isothermal elliptical $\beta$ model. The second samples includes 38
galaxy clusters~\cite{Boname06},  whose plasma and matter
distributions  were analyzed by assuming hydrostatic equilibrium
model and spherical symmetry. Therefore, the CDDR tests are sensitive to the models for the cluster gas distribution, since the ADD data is closely relate to the assumption of cluster models. For the galaxy cluster samples, the
statistical   and systematic errors account for about $20\%$ and
$24\%$~\citep{Reese20023,Boname06} and
they are combined in quadrature for the
ADD~\cite{holanda20103,Boname06}.

The observational ADD data  can  be also obtained from the BAO measurements~\cite{Bassett2010}. The early universe consists of a hot, dense plasma of electrons and baryons. Photon is coupled with the baryons via Thomson scattering.  A system of standing sound waves within the plasma can be created on account of the existence of a competition between radiation pressure and gravity, so called BAOs. As the universe expands, the plasma cools to below $3000K$---a low enough energy such that the electrons and protons in the plasma could combine to form neutral hydrogen atoms, i.e. recombination. The free electrons are quickly captured and the coupling between photons and baryons ends abruptly, which leads to a overdensity of baryons at the scale about 150Mpc today. This scale can be observed in the clustering distribution of galaxies today and can be used as a standard ruler. One can obtain the ADD  by
 combining of the measurements of the baryon acoustic peak and the Alcock-Paczynski distortion from galaxy clustering (see~\cite{Bassett2010} for a review).  The five  ADD data points from BAO measurements are released by the WiggleZ Dark
Energy Survey~\cite{Blake2012}, the Sloan Digital Sky Survey
(SDSS)~\cite{Xu2013} and Data Release 11~\cite{Samushia2014} ( also listed in Table I from Ref.~\cite{Wu2015}).

The ET is the third generation of the ground based GW detector, and it, as proposed by programme, consists of three collocated underground arms with the length of 10 km  and a $60^\circ$ opening  angle. The ET  would be able to detect GW signals to be ten times more sensitive in amplitude than the advanced ground-based detectors, covering a wide frequency range of $1\sim 10^4$ Hz with the redshits range  $z\sim 2$ for the NS-NS  and $z> 2$ for the BH-NS mergers systems. If compact mergers are NS-NS or BH-NS binaries, the source redshift may be observed from EM counterparts that occur coincidentally with the GW events~\cite{Zhao2011,Nissanke2010,Cai2017}. Thus, the LD-redshifts  relation can be constructed with a cosmological-model-independent way, and it can be employed to make constraints on the basic parameters of cosmology.   The ratio between NS-NS and BH-NS binaries, in this work, is taken to be 0.03, as illustrated by the Advanced LIGO-Virgo network~\cite{Abadie2010a}. Here, for brevity, we only summarize the process of Refs.~\cite{Zhao2011,Cai2017,Wei2018,Yang2017} in which  observations of GW from future ET are simulated, and then we will forecast the constraints on CDDR.

\begin{figure}[htbp]
\includegraphics[width=8cm]{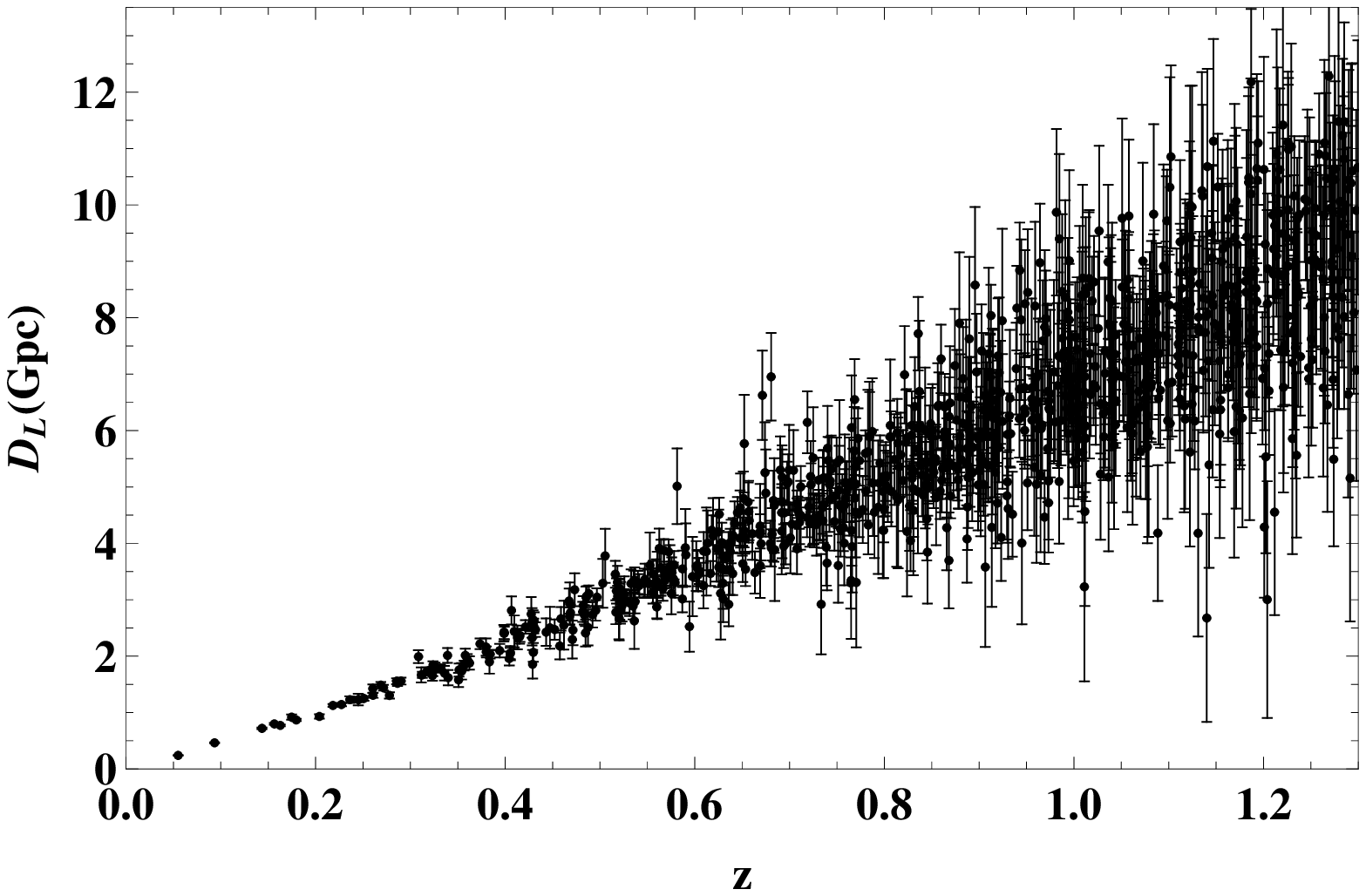}
\includegraphics[width=8cm]{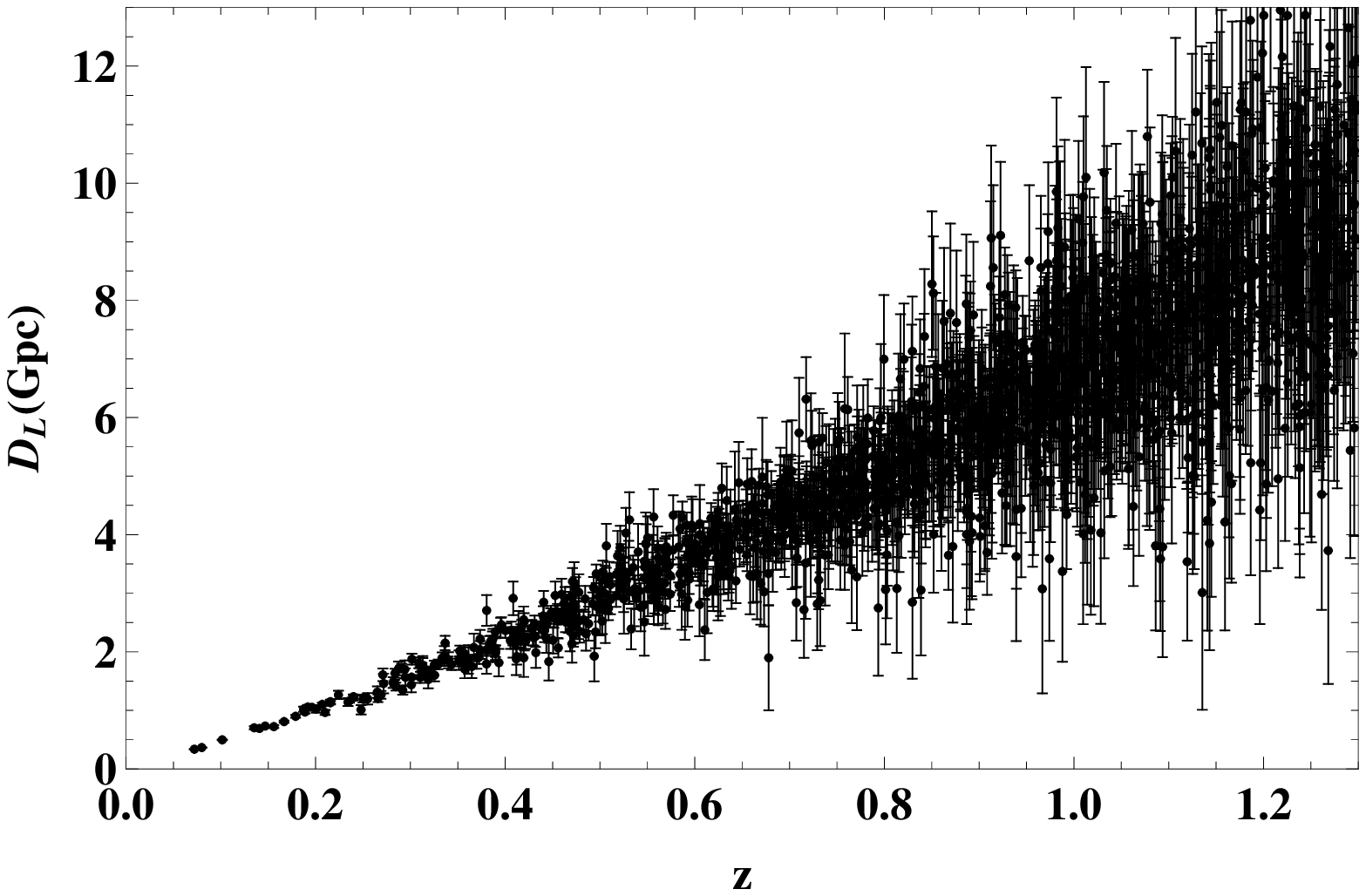}
\caption{\label{mock1} The example catalogues with 930 (left) and 1700 (right) observed GW events of redshifts, LD and the error of LD from the fiducial model in the redshifts region $0<z<1.3$, in which 550 and 1000 data points are located in the redshift region $0<z<1$, respectively. }
\end{figure}

For the waveform of GW, the stationary phase approximation is applied to compute the Fourier transform $\mathcal{H}(f)$ of the time domain waveform $h(t)$,
\begin{align}
\mathcal{H}(f)=\mathcal{A}f^{-7/6}\exp[i(2\pi ft_0-\pi/4+2\psi(f/2)-\varphi_{(2.0)})],
\label{equa:hf}
\end{align}
where the Fourier amplitude $\mathcal{A}$ is given by
\begin{align}
\mathcal{A}=&~~\frac{1}{D_{\rm L}}\sqrt{F_+^2(1+\cos^2(\iota))^2+4F_\times^2\cos^2(\iota)}\nonumber\\
            &~~\times \sqrt{5\pi/96}\pi^{-7/6}\mathcal{M}_{\rm c}^{5/6}\,,
\label{equa:A}
\end{align}
where $\mathcal{M}_{\rm c}$ denotes the observational chirp mass and ${D_{\rm L}}$ is the LD which plays the most important role in this test. The definition of the beam pattern functions $F_{+,\times}$, the polarization angle $\psi$, the epoch of the merger $t_0$, the phase parameter such as the angle of inclination $\iota$ and $\varphi_{(2.0)}$ are introduced  in Refs.~\cite{Zhao2011,Cai2017,Wei2018,Yang2017}.   The cosmological parameters of fiducial concordant  model are adopted with the most recent Planck results~\cite{Plank2015}:
\begin{align}
h_0=0.678,~~~\Omega_{\rm m}=0.308,~~~\Omega_{\rm k}=0,~~~w=-1,
\label{equa:A1}
\end{align}
where $H_0=100 h_0{\rm km s^{-1}Mpc^{-1}}$, $\Omega_{\rm m},\Omega_{\rm k}$ and $w$ denote the Hubble constant,
dark matter density parameter, the cosmic curvature parameter today and the dark energy equation of state respectively. It is known that the redshift range of ADD data from galaxy cluster and BAO  is in the region $0<z<1$, and the corresponding number of data points from SNIa Union 2 or Union 2.1 compilation is 537 or 551, respectively. In order to compare our results relevantly with  the number density of data points from Union compilation in this redshifts region, we first  simulate 550 data points (set A) from future GW events. We also simulate 1000 data points (set B) to study the impact of the quantity  of GW data on the test. The mock  results are shown in Fig.~(\ref{mock1}).

\section{ Methods}
The most straightforward method to test CDDR is to compare the LD with the ADD at the same redshifts through the identity of Eq.~(\ref{ddr}). Generally, in the checking process, some departures from CDDR are allowed  through defining the following function
\begin{equation}\label{PCDD}
  {D_{\rm L}(1+z)^{-2}\over D_{\rm A}}=\eta{(z)}\,.
\end{equation}
 The CDDR holds while   $\eta(z)=1$. All deviations from CDDR, which occur possibly at some redshifts, will be encoded in the function $\eta{(z)}$.
 In this work, four potential parameterizations for the $\eta{(z)}$ are adopted, namely,  linear one $\eta(z)=1+\eta_0z$ ($\rm P_1$), and three non-linear ones,  $\eta(z)=1+\eta_0z/(1+z)$ ($\rm P_2$), $\eta(z)=1+\eta_0z/(1+z)^2$ ($\rm P_3$), and $\eta(z)=1+\eta_0\ln(1+z)$ ($\rm P_4$).

In principle, given a  ADD sample from each galaxy cluster or BAO system, one should select a  LD ($D_{\rm L}(z)$) data point
 from GW data that shares the same redshift $z$ with the given ADD data
 to test the CDDR. However, this condition usually can not be met in
 recent astronomical observations.
 To achieve this goal, a number of  methods have been proposed~\cite{holanda20103,Li2011,Liao2016,Meng2012,Yang2017}. In order to compare our results with that from Refs.~\cite{Meng2012,Wu2015}, we employ a cosmological model-independent binning methods to obtain the LD ($D_{\rm L}(z)$) from certain GW data points.

\subsection{Method: Binning the GW data}
In order to test the validity of CDDR with a cosmological-model-independent way, Holanda {\it et al.}
~\cite{holanda2010,holanda20103}, Li {\it et al.}~\cite{Li2011} and Liao {\it et al.}~\cite{Liao2016} adopted a selection criterion
$\Delta z=|z_{\rm ADD}-z_{\rm SNIa}|<0.005$, where $z_{\rm ADD}$ and $z_{\rm SNIa}$ denote the redshift of a ADD sample and  SNIa data respectively, and chose the nearest SNIa data
to match a ADD sample. However, using merely one SNIa data point from all
those available which meet the selection criterion will lead to larger statistical errors.
In order to avoid them, instead of using the nearest point of Union 2.1 SNIa,
Wu {\it et al.}~\cite{Wu2015},  Meng {\it et al.}~\cite{Meng2012} bin these data available to obtain LD to match the corresponding ADD sample. Following the process, we bin the simulated GW data which meet the criterion. In order to avoid correlations among the individual CDDR tests, we choose the LD samples  with a procedure that the data points will not be used again if they have been matched to some cluster or BAO samples.  In this method, we employ an inverse variance
weighted average of all the selected data. If $D_{{\rm L}i}$ denotes the
$i$th appropriate luminosity distance data points with $\sigma_{D_{{\rm L}i}}$
 representing the corresponding observational uncertainty,  we
 can straightly  obtain the following  with conventional
 data reduction techniques in Chapter.($4$) of the Ref.~\cite{Bevington2003},
\begin{equation}
\label{avdi1}
\bar{D_{\rm L}}={\sum(D_{{\rm L}i}/\sigma_{D_{{\rm L}i}}^2)\over \sum1/\sigma_{D_{{\rm L}i}}^2},
\end{equation}
\begin{equation}
\label{erroravdi1}
\sigma^2_{\bar{D_{\rm L}}}={1\over \sum1/\sigma_{D_{{\rm L}i}}^2},
\end{equation}
where $\bar{D_{\rm L}}$ represents the weighted mean luminosity distance
and $\sigma_{\bar{D_{\rm L}}}$ corresponds its uncertainty.

\begin{figure}[htbp]
\includegraphics[width=7.5cm]{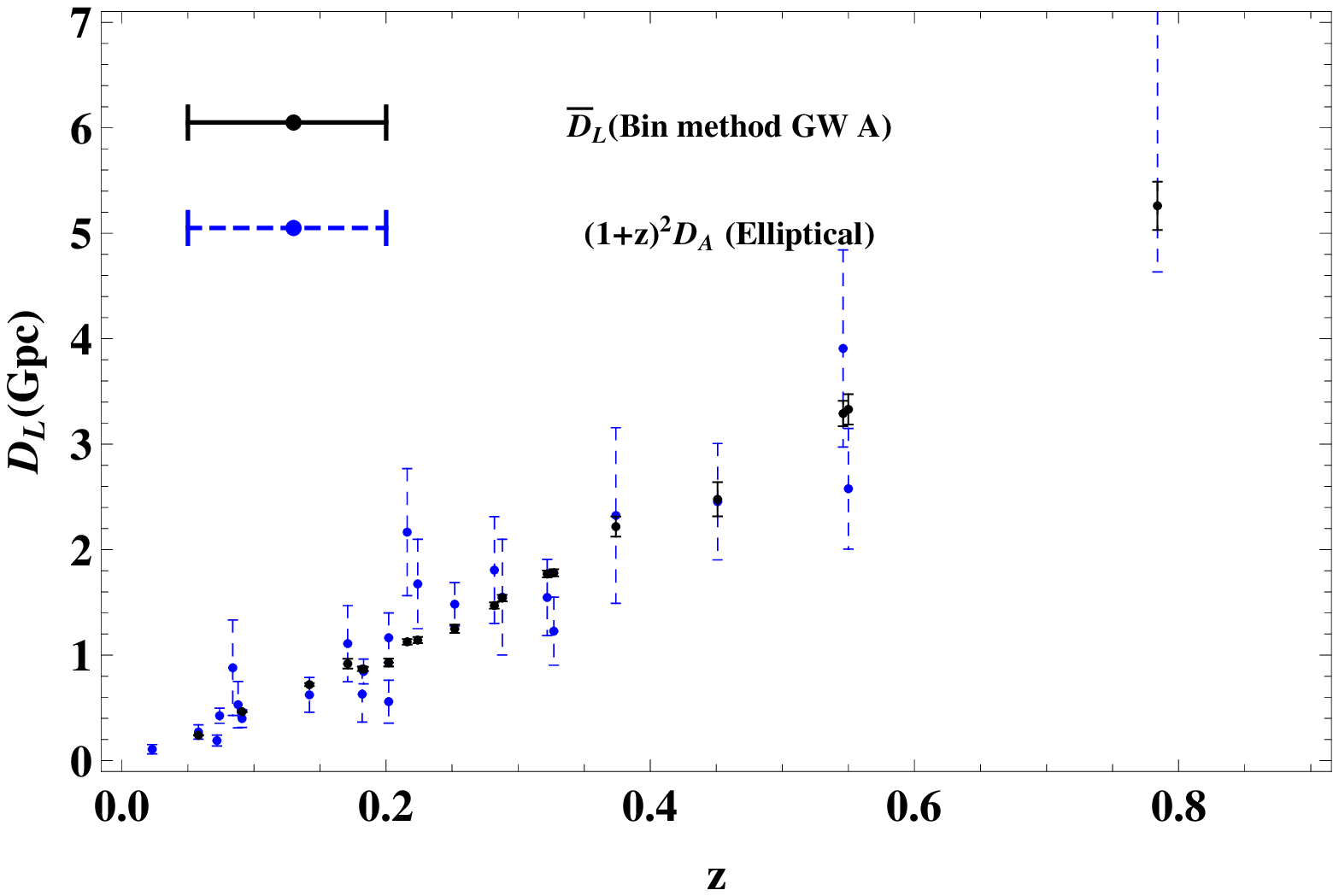}
\includegraphics[width=7.5cm]{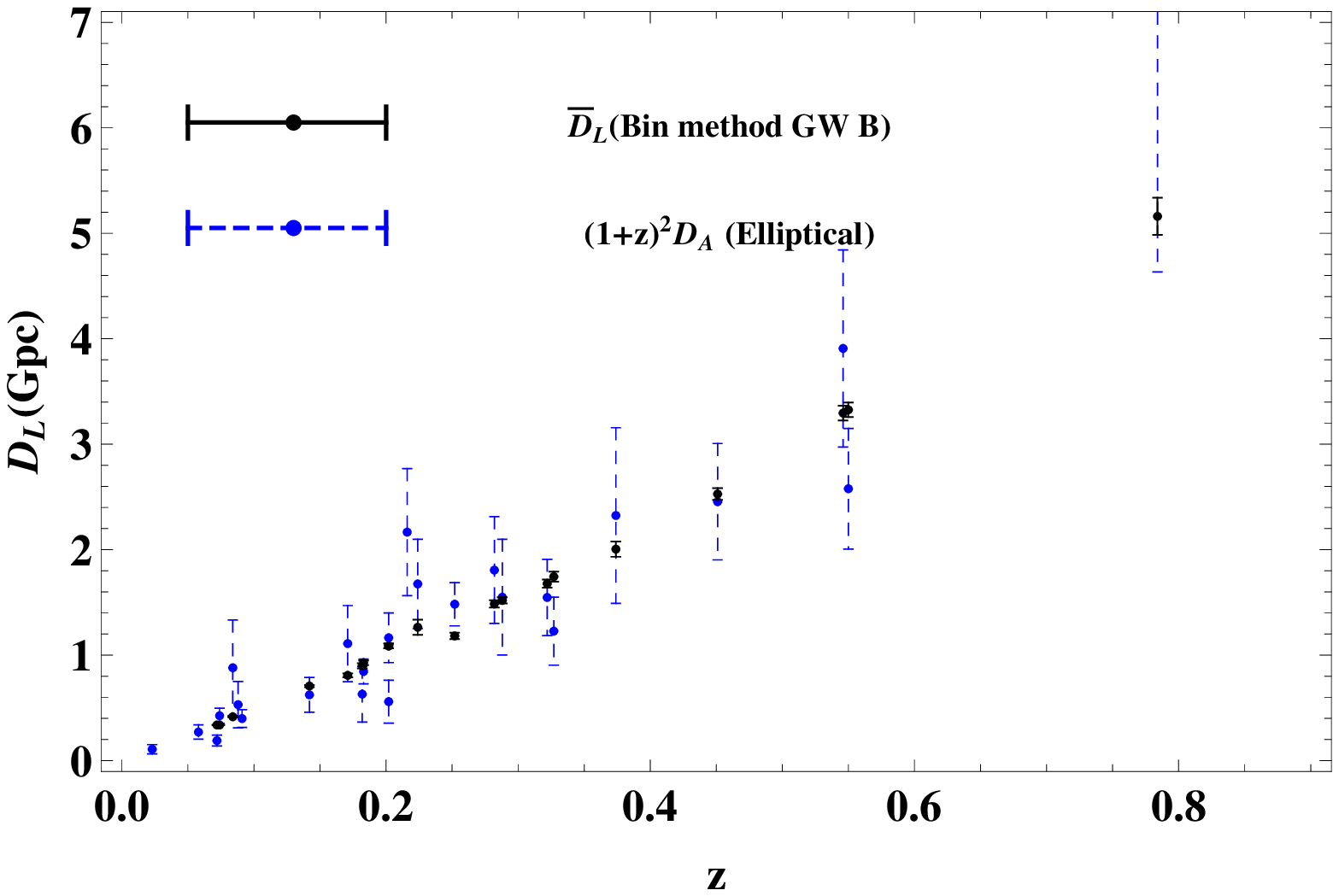}
\includegraphics[width=7.5cm]{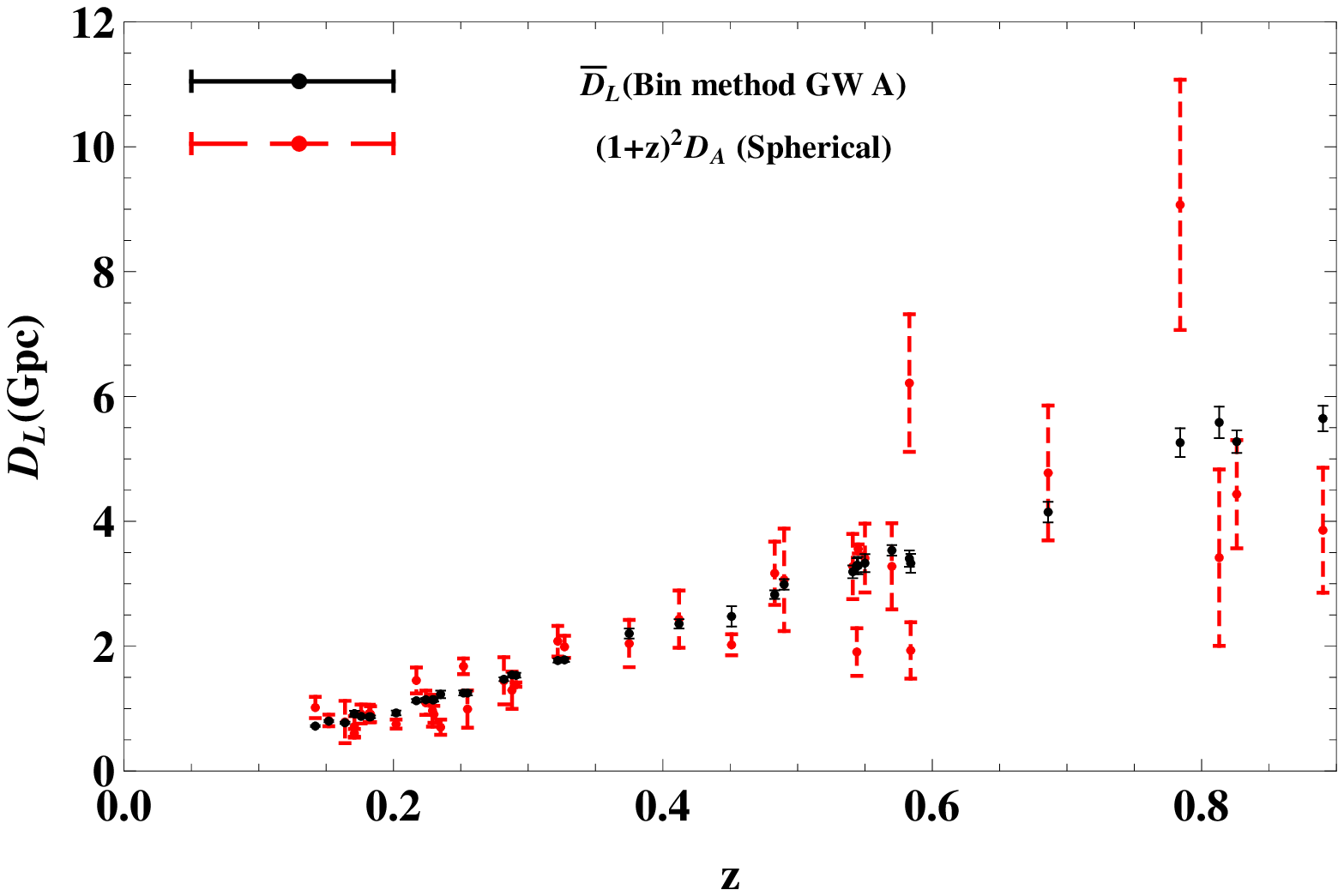}
\includegraphics[width=7.5cm]{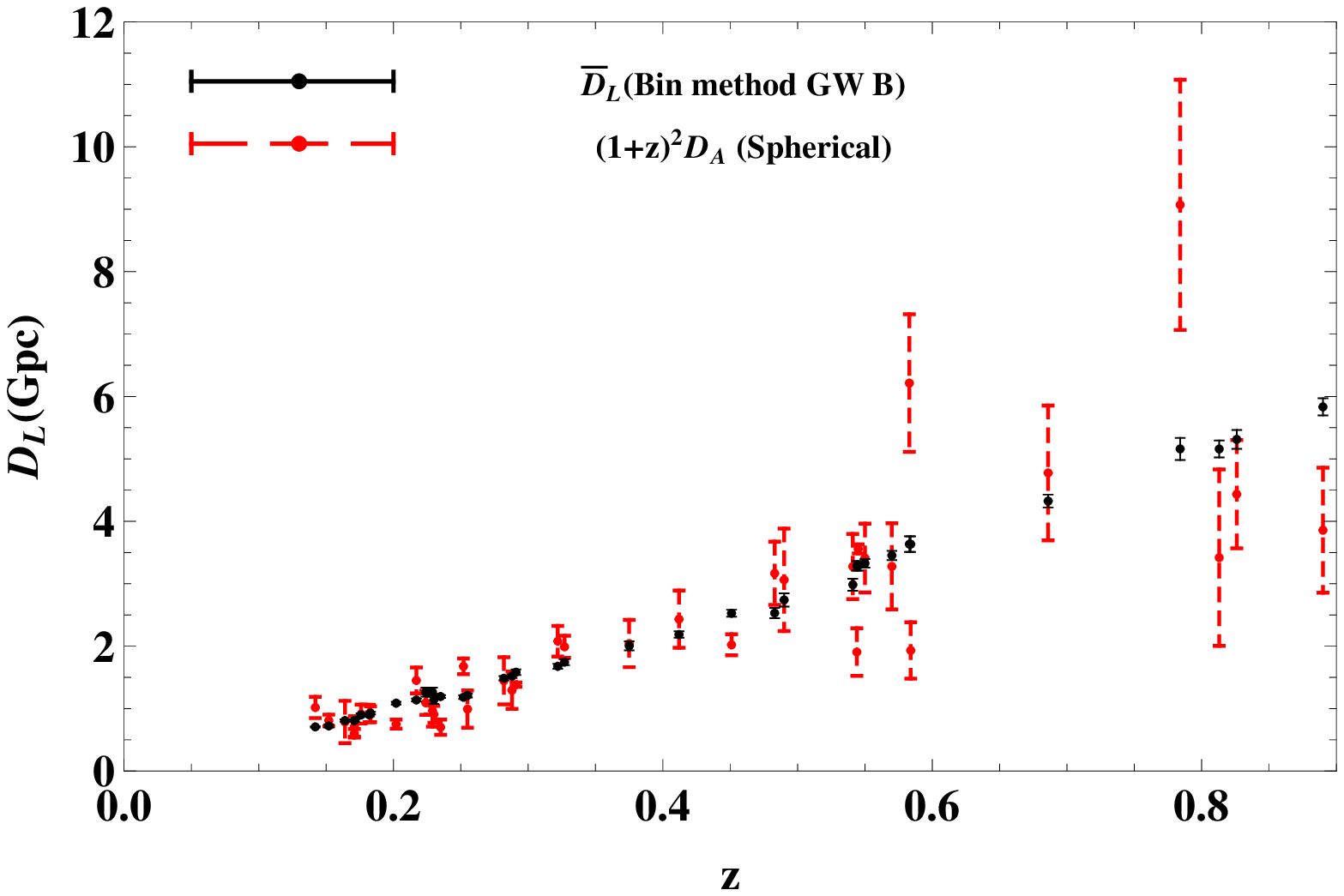}
\includegraphics[width=7.5cm]{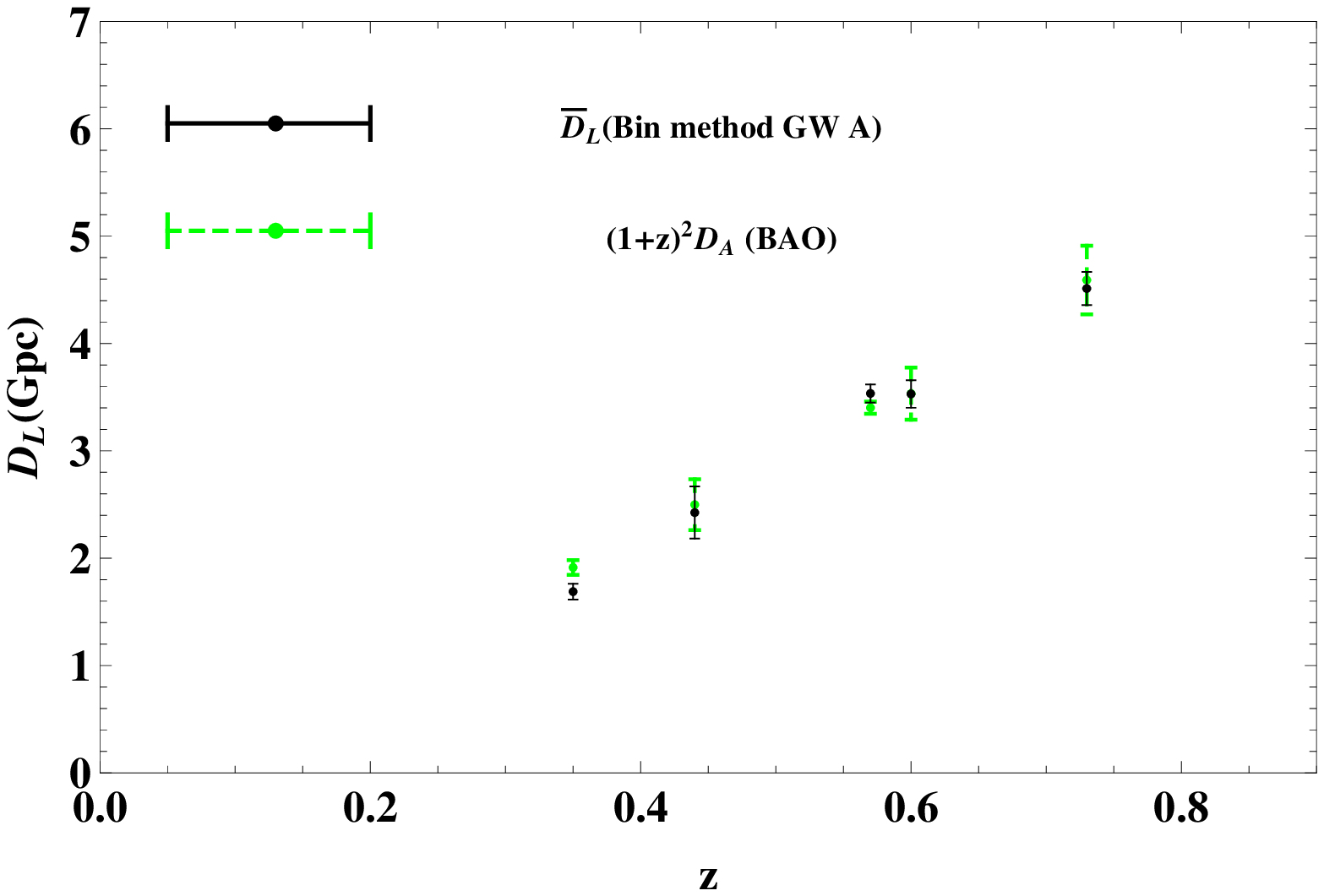}
\includegraphics[width=7.5cm]{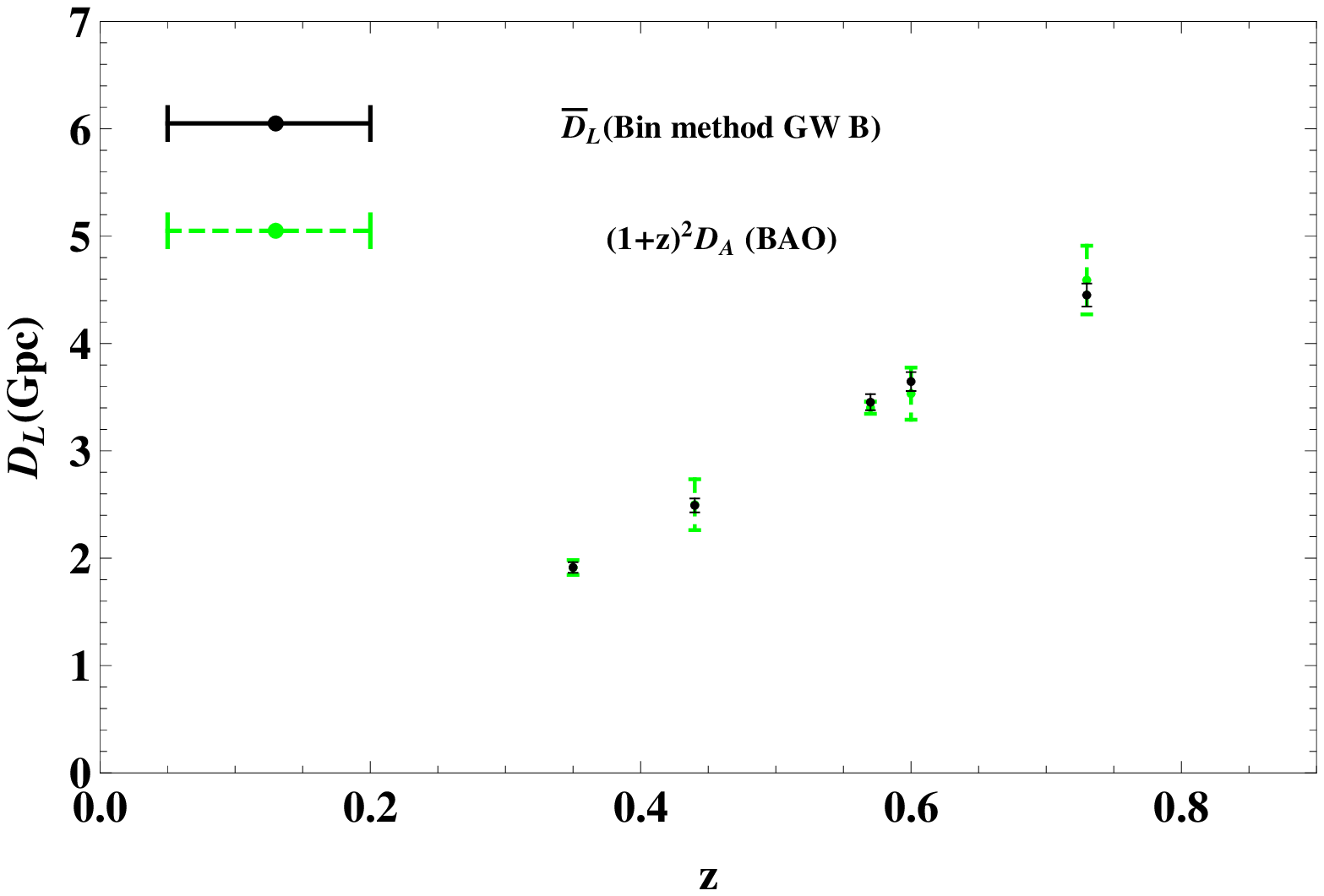}
\caption{\label{fighubble} The distribution of ADD samples from galaxy cluster of elliptical $\beta$ model, spherical $\beta$ model and  BAO corresponding to the LD obtained with this method from set A and set B of simulated GW data. The samples of elliptical $\beta$ model without being coupled with LD data points in the redshift range $z<0.1$  do not satisfy the selection criteria, and they should be discarded in the analysis. }
\end{figure}

 The selection criterion
 can be generally satisfied with all  the samples of  spherical $\beta$ model and  BAO. However,   for the elliptical $\beta$ model, only 20 and 21 samples  satisfied with this selection criteria for set A and B from the simulated GW data except for some data
points in the low redshift region $z<0.1$, since the number density of the simulated GW data in this region is very small.
The distributions of ADD samples and LD data obtained by this method are shown in Fig.(~\ref{fighubble}).

\section{Analysis and Results}
\begin{figure}[htbp]
\includegraphics[width=7.5cm]{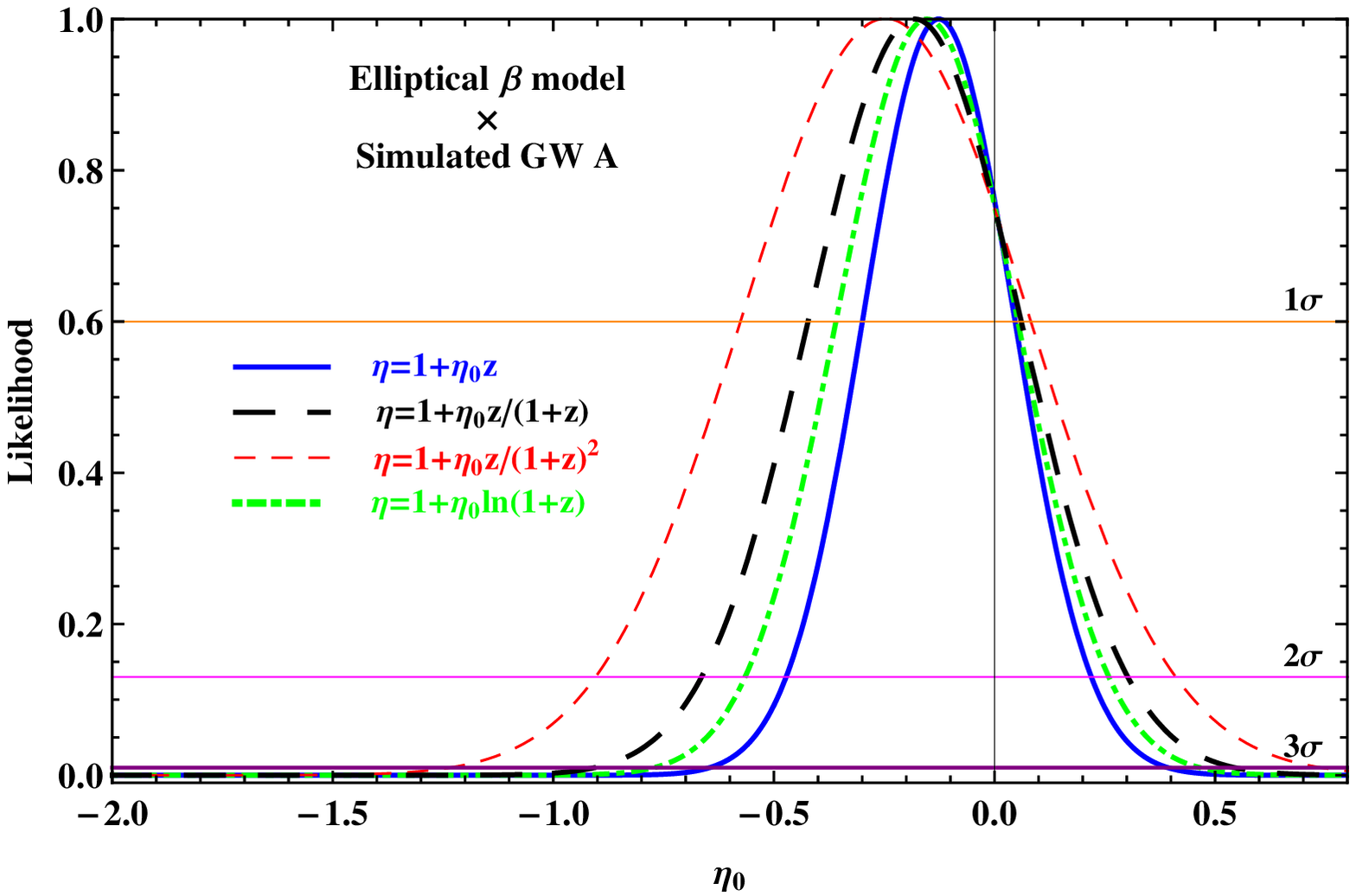}
\includegraphics[width=7.5cm]{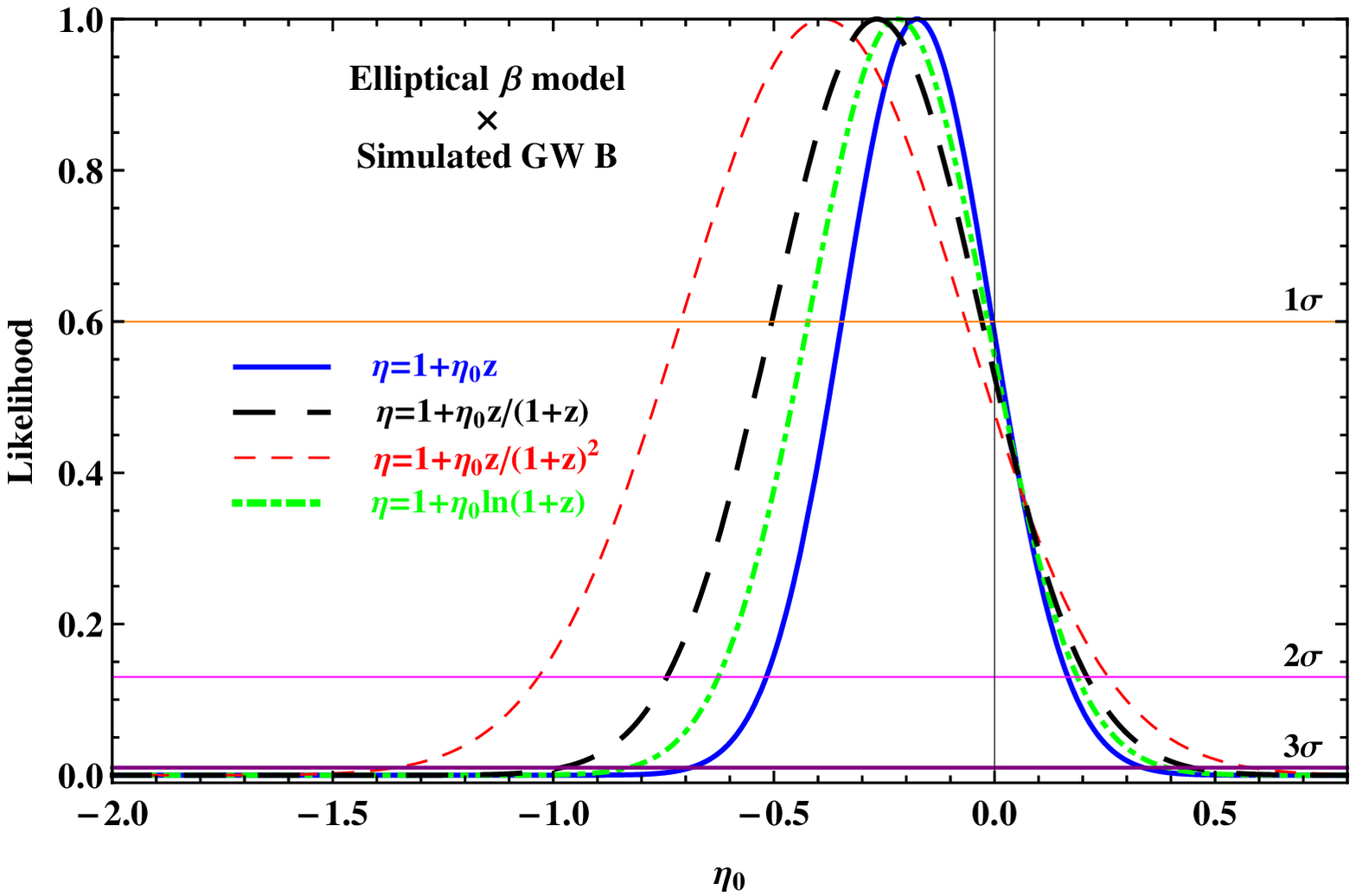}
\includegraphics[width=7.5cm]{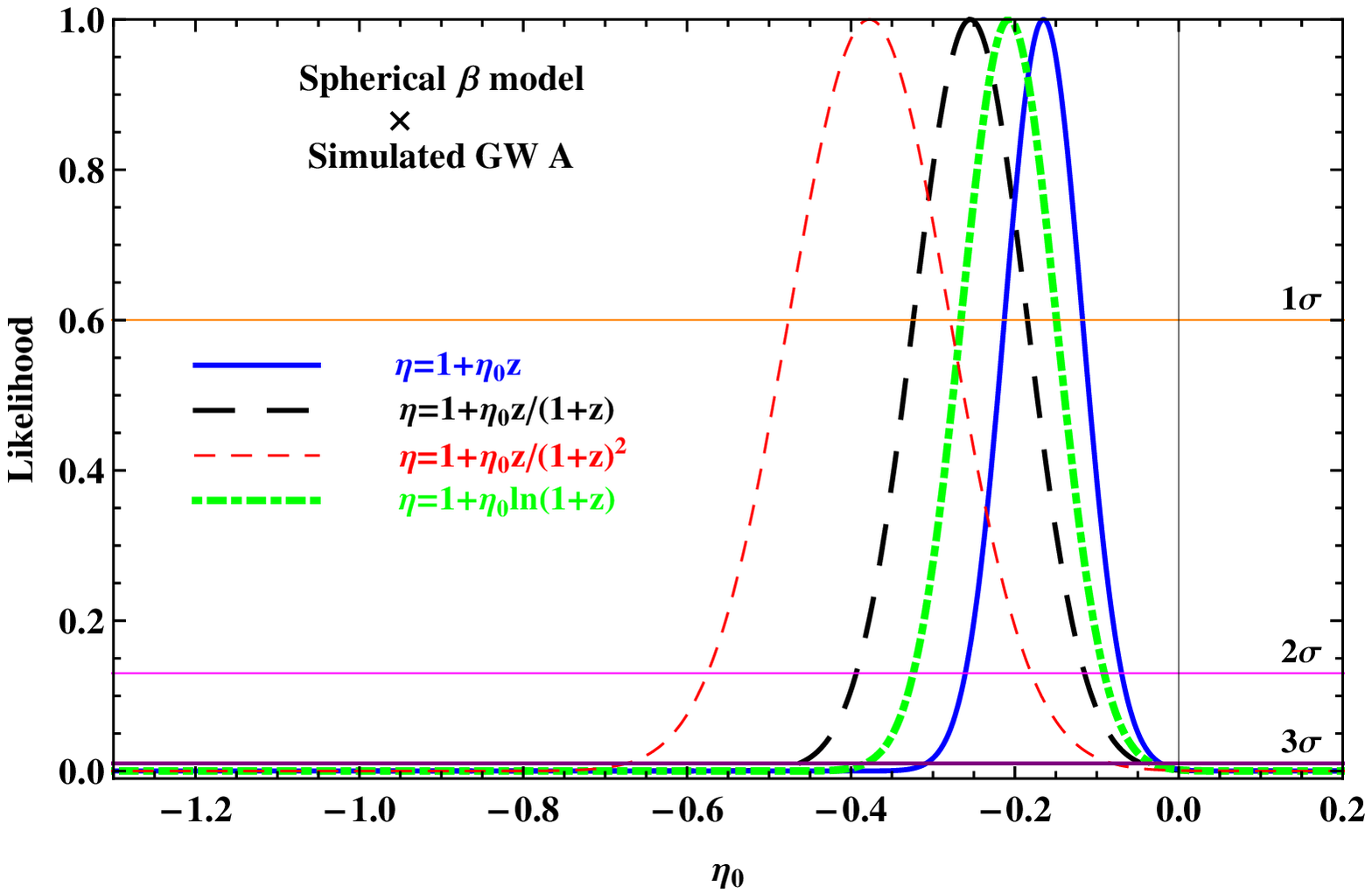}
\includegraphics[width=7.5cm]{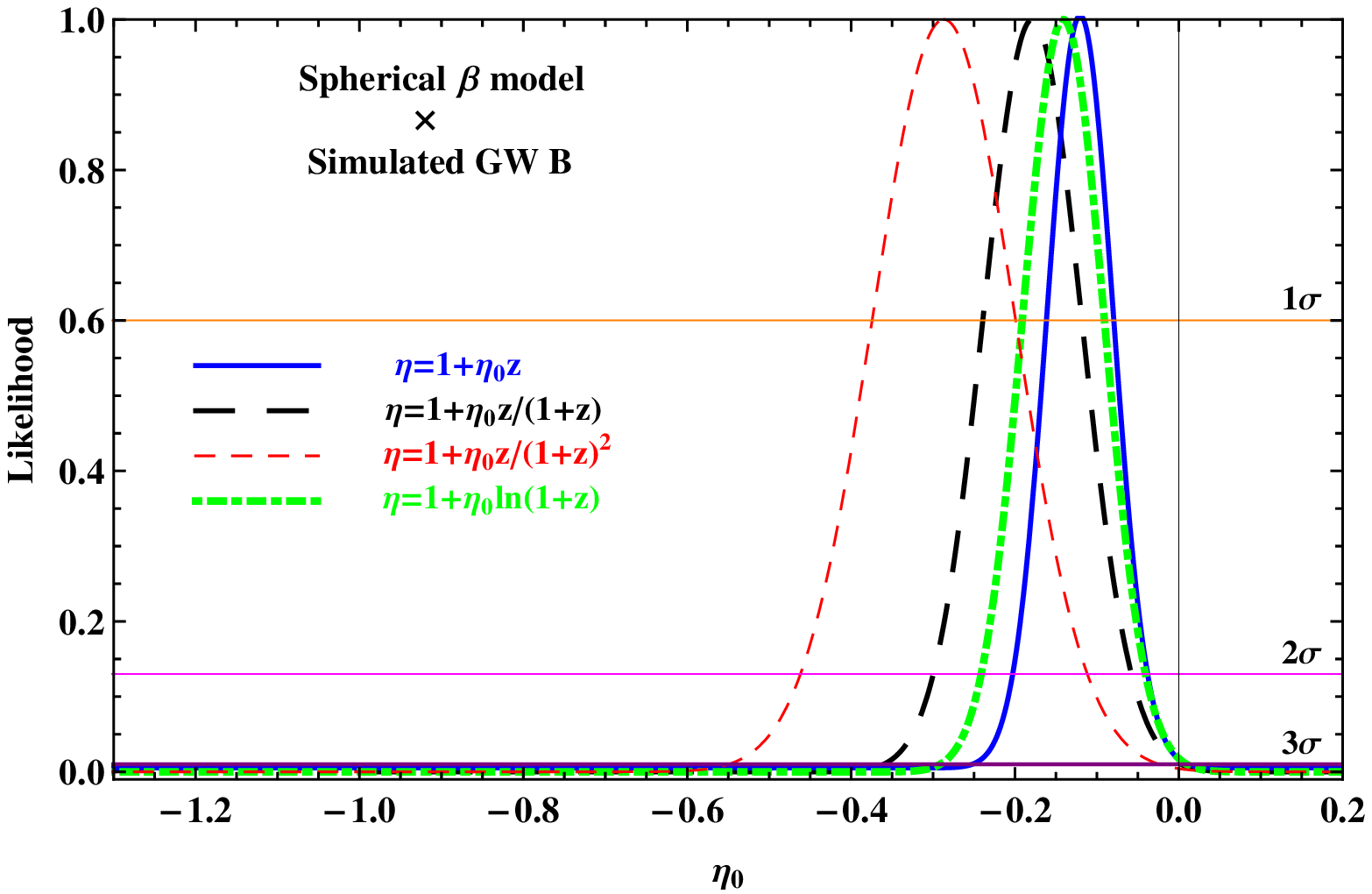}
\includegraphics[width=7.5cm]{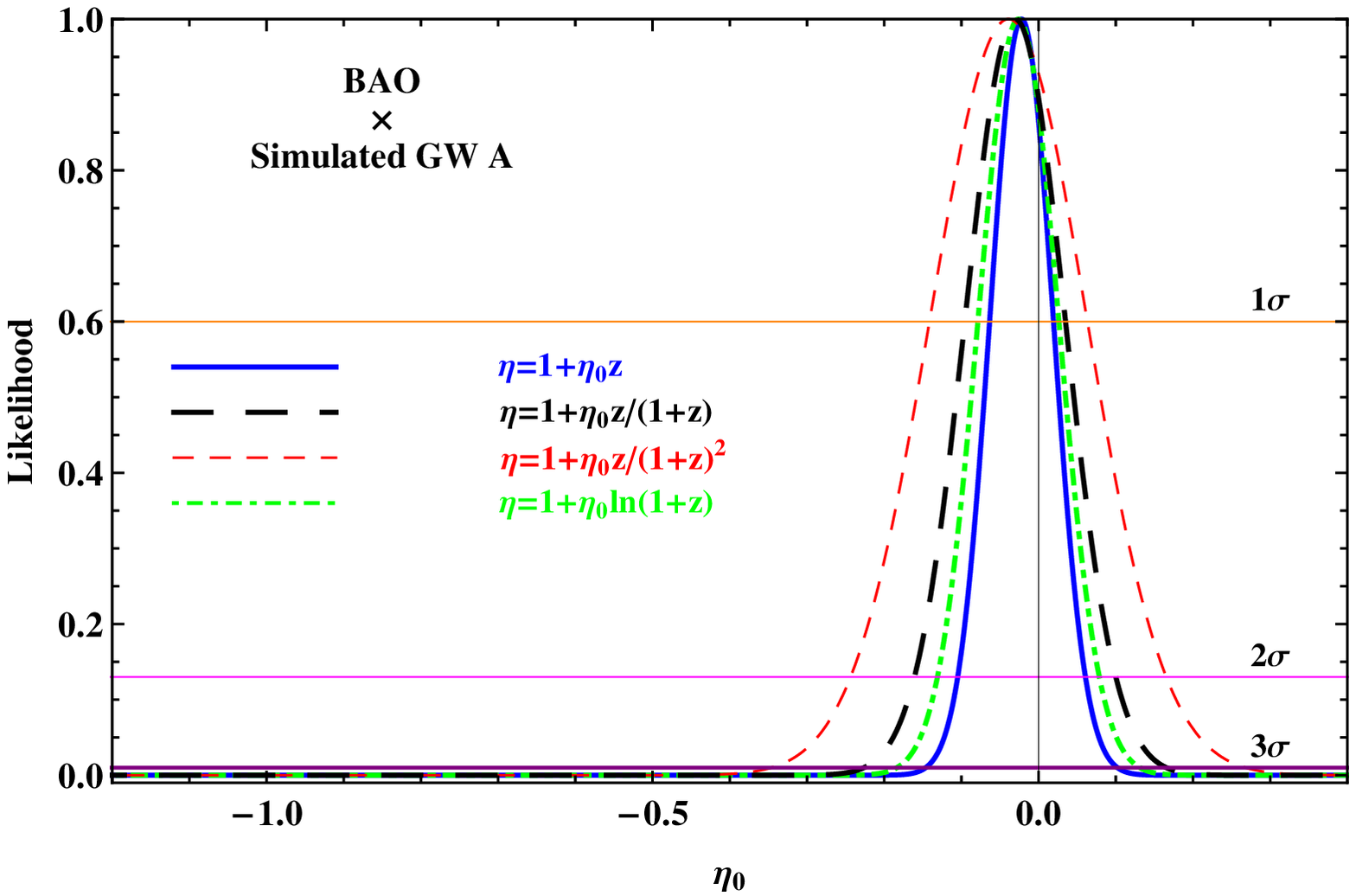}
\includegraphics[width=7.5cm]{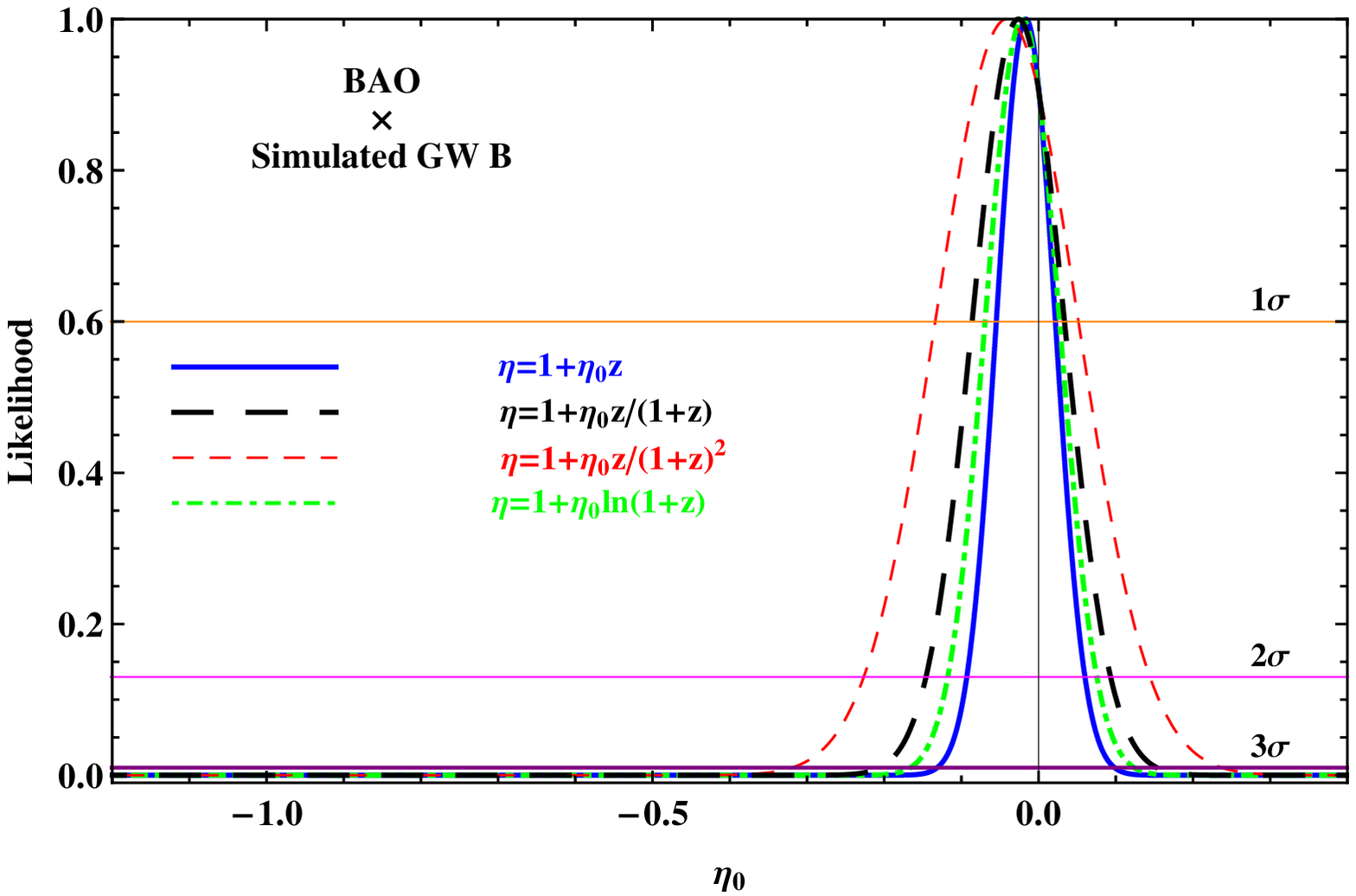}
\caption{\label{Figlikec} The likelihood  distribution functions obtained from set A and set B of the simulated GW data points. }
\end{figure}

To place constraints on $\eta_0$, one must firstly obtain
$\eta_{\rm obs}(z)$ with
\begin{equation}
\label{eta}
\eta_{\rm obs}(z)=\bar{D_{\rm L}}(z)(1+z)^{-2}/D_{\rm A}(z)
\end{equation}
from
the  galaxy cluster or BAO samples and  the luminosity distance from the binning method. The corresponding error of $\eta_{\rm obs}$ can be obtained through
\begin{equation}
\label{chi3sigma}
\sigma^2_{\eta_{\rm obs}}={\eta_{\rm obs}^2}\left[\left({\sigma_{D_{\rm A}(z)}\over{D_{\rm A}(z)}}\right)^2+\left(\sigma_{D_{\rm L}(z)}\over{D_{\rm L}(z)}\right)^2\right].
\end{equation}
Thus, using
 the following equation
\begin{equation}
\label{chi3} \chi^{2} = \sum_{i}\frac{{\left[\eta(z)-
\eta_{obs}(z) \right]^{2} }}{\sigma^2_{\eta_{obs}}}\,,
\end{equation}
one can obtain the constraints on $\eta_0$.
Results are shown in Fig.~(\ref{Figlikec}) and Tab.~(\ref{likelihood1}).

\begin{table}[htp]
\begin{tabular}{|c|c|c|c|c|c|c|}
\hline
\scriptsize{Parametrization }  & \   $\eta_0^{\rm {A}}$\tiny(elliptical)\ \   &$\eta_0^{\rm B}$\tiny(elliptical)& \ \  $\eta_0^{\rm A}$ \tiny(spherical)\ \ \   &$\eta_0^{\rm B}$ \tiny(spherical)& \ \  $\eta_0^{\rm A}$\tiny(BAO) \ \ \   &$\eta_0^{\rm B}$\tiny(BAO)\\
\hline
\scriptsize{$1+\eta_0 z$ } &  \scriptsize{${-0.126{\pm{0.173}}}$} & \scriptsize{${-0.175{\pm{0.171}}}$}&  \scriptsize{${-0.165{\pm{0.048}}}$} & \scriptsize{${-0.113{\pm{0.041}}}$}
&\scriptsize{${-0.022{\pm{0.041}}}$ }& \scriptsize{${-0.016{\pm{0.037}}}$} \\
\hline
\scriptsize{$1+\eta_0 {z\over 1+z}$} & \scriptsize{${-0.180{\pm{0.242}}}$}&\scriptsize{ ${-0.267{\pm{0.238}}}$}  &\scriptsize{${-0.254{\pm{0.070}}}$}& \scriptsize{${-0.178{\pm{0.060}}}$}
&\scriptsize{${-0.030{\pm{0.065}}}$} & \scriptsize{${-0.026{\pm{0.060}}}$} \\
\hline
\scriptsize{$1+\eta_0 {z\over (1+z)^2}$} & \scriptsize{${-0.247{\pm{0.328}}}$}& \scriptsize{${-0.388{\pm{0.323}}}$}
 &\scriptsize{${-0.378{\pm{0.100}}}$} & \scriptsize{${-0.287{\pm{0.087}}}$}
 &\scriptsize{${-0.043{\pm{0.101}}}$} & \scriptsize{${-0.041{\pm{0.092}}}$} \\
\hline
\scriptsize{$1+\eta_0 {\ln(1+z)}$} & \scriptsize{${-0.153{\pm{0.206}}}$} & \scriptsize{${-0.219{\pm{0.203}}}$} &  \scriptsize{${-0.207{\pm{0.058}}}$} & \scriptsize{${-0.141{\pm{0.050}}}$}
&  \scriptsize{${-0.026{\pm{0.052}}}$} & \scriptsize{${-0.021{\pm{0.048}}}$}  \\
\hline
\end{tabular}
\caption{The summary of maximum likelihood estimation results of $\eta_0$ for four parameterizations respectively. The $\eta_0$ is represented by the best fit value at 1 $\sigma$ CL for each data set. The superscript A and B  represents the case obtained from set A  and set B of the   simulated GW data, respectively.}
\label{likelihood1}
\end{table}

For the galaxy cluster, seen from Fig.~(\ref{Figlikec}) and Tab.~(\ref{likelihood1}), it can be obtained that the CDDR is consistent with the elliptical $\beta$ model and the simulated GW data set A or B at $1\sigma$ CL or  $2\sigma$ CL respectively, which is consistent with that from refs.~\cite{holanda20103,Li2011,Meng2012}. However, for the spherical $\beta$ model, the CDDR is not compatible with the observational data even at $3\sigma$ CL. This result suggests a stronger violation than that from Refs.~\cite{holanda20103,Li2011,Meng2012}.   It should be noted that, unlike SNIa observations, the violation from the CDDR is obtained from the LD of GW measurements which is insensitive to the conservation of photon number. If photon does follow along the null geodesics in a Riemannian geometry, the CDDR is valid  in the test from GW measurements. One may conclude that the violation of the CDDR may be resulted from the existence of large deviation while the spherical $\beta$ model is used to describe the structure of galaxy cluster. For BAO samples, the CDDR is compatible with the observational data at $1\sigma$ CL, which is consistent with the results from Ref.~\cite{Wu2015}. As for the four parameterizations, one can conclude that the linear form can provide the strictest constraints  and the best fits on the test for the CDDR.

By comparing the constraints on the $\eta_0$ at $1\sigma$ CL, for the spherical $\beta$ model, we obtain error bars about $50\%$ smaller than that obtained from Ref.~\cite{Meng2012} where Union 2 SNIa compilation are used, regardless the $\eta(z)$ functions adopted. For the BAO samples, the results are about $35\%$ smaller than that from Ref.~\cite{Wu2015}, where $\eta_0={-0.027{\pm{0.064}}}$ and ${-0.039{\pm{0.099}}}$ with  $\rm P_1$ and $\rm P_2$ of the $\eta(z)$ functions,  respectively. For the elliptical $\beta$ model, the error bars are similar to that from Ref.~\cite{Meng2012}. However, it should be noted that the number of available samples is only 20, which is less than that used in Ref.~\cite{Meng2012}. As one may see that,  much tighter constraints can be obtained by using  future measures of GW events while the same number of ADD samples are considered. By comparing the results from simulated GW data set A with the results from set B at $1\sigma$ CL, one may find that the tests are almost independent on the quantity of simulated GW data, which may show that the errors of galaxy clusters and BAO samples dominate the constraints on the CDDR.

\section{conclusion and discussion}
The cosmic distance-duality  relation (CDDR) plays a fundamental role in astronomical observations and modern cosmology. Its  validation with various observational data  is an important issue in modern cosmology, as any  violation of it could be a signal of new physics in the modern theory of gravity or in particle physics. However, most of the previous tests involving the luminosity distance from SNIa on the CDDR are sensitive to the conservation of number of photons.

The direct detections of the gravitation wave (GW) events  have thrown the  observational cosmology into a new era of multi-messenger.  More precisely, for this astronomical observation, the luminosity distances (LD) can be measured from the waveform and amplitude of the gravitational wave, and it is insensitive to a possible non-conservation of the number of photons.

In this work, we have simulated 550 and 1000 data points of future GW measurements from the Einstein Telescope. The angular diameter distances (ADDs) are from  the galaxy clusters samples~\cite{DeFilippis05,Boname06} obtained by using SZE and X-ray surface brightness observations  and the BAO data~\cite{Wu2015}.  In order to compare our results to that from Refs.~\cite{Meng2012,Wu2015} where Union 2 or Union 2.1 SNIa are adopted,  550 data points are adopted to ensure that the average number density of the  GWs is nearly equal to the number density  of the SNIa Union compilation in this redshift range, and  the binning method is employed to obtain the corresponding LDs from simulated GW data for each  BAO or galaxy cluster system. Then we  detect the potentialities of future GW measurements to test the CDDR. The results show that  future GW measurements may  provide much tighter  constraints on the CDDR,  while we compare the results at $1\sigma$ CL with previous ones from SNIa Union2.1 or Union2~\cite{Meng2012,Wu2015} if the same number of ADD samples are adopted. With the increase of the measuring quality and quantity of future observations, we can forecast that future GW measurement can be considered as a powerful tool to validate this reciprocal relation.

\begin{acknowledgments}

We very much appreciate helpful comments and suggestions from anonymous referees, and also like to thank for helpful discussion from Puxun Wu and Zhengxiang Li. This work was supported by the National Natural Science Foundation of China
under Grant Nos. 11147011, the
Hunan Provincial Natural Science Foundation of China under Grant No. 12JJA001, the National Natural
Science Foundation of China under Grants Nos. 11465011,11865011,
the Foundation of Department of science and technology of Guizhou
Province of China under Grants No. J [2014] 2150 and the Foundation of
the Guizhou Provincial Education Department of China under Grants
No. KY [2016]104.

\end{acknowledgments}


\begin{thebibliography}{99}
\bibitem{eth1933}  I. M. H. Etherington,  Phil. Mag. {\bf 15}, 761 (1933);
         reprinted in  Gen. Relativ. Gravit. {\bf 39}, 1055 (2007).

\bibitem{ellis1971}G. F. R. Ellis,
         Proc. Int. School Phys. Enrico Fermi, R. K. Sachs (ed.),
         pp. 104-182 1971(Academic Press: New York) reprinted in  Gen.\ Rel.\
         Grav. {\bf 41}, 581 (2009).

\bibitem{ellis2007} G. F. R. Ellis,  Gen. Relativ. Gravit. {\bf 39}, 1047 (2007).
\bibitem{uzan2004} J. P. Uzan, N. Aghanim and Y. Mellier,
                Phys. Rev. D {\bf 70}, 083533 (2004). [astro-ph/0405620].
\bibitem{Santana2017} L. T. Santana, M. O. Calvao, R. R. R. Reis and B. B. Sif
        fert,  Phys. Rev. D 95, no. {\bf 6}, 061501 (2017).
        [arXiv:1703.10871 [gr-qc]].

\bibitem{Csaki2002} C. Csaki, N. Kaloper and J. Terning, Phys. Rev. Lett. {\bf 88}, 161302 (2002).
\bibitem{Lima2011} J. A. S. Lima, J. V. Cunha and V. T. Zanchin,  \apj 742, 2(2011)  [arXiv:1110.5065].

\bibitem{Hees2014} A. Hees, O. Minazzoli and J. Larena,  Phys. Rev. D {\bf 90}, 124064 (2014)
              [arXiv:1406.6187 [astro-ph.CO]].

\bibitem{Holanda20161} R. F. L. Holanda and S. H. Pereira,  Phys. Rev.
            D {\bf 94}, no. 10, 104037 (2016) [arXiv:1610.01512 [astro-
          ph.CO]].
\bibitem{Holanda2017} R. F. L. Holanda, S. H. Pereira and  S. Santos-da-Costa, Phys. Rev. D {\bf 95}, no. 8, 084006 (2017)
              [arXiv:1612.09365 [astro-ph.CO]].
\bibitem{Aguirre1999} A. Aguirre, \apj {\bf 525}, 583 (1999).


\bibitem{Avgoustidis2009} A. Avgoustidis, L. Verde and R. Jimenez, J. Cosmol. Astropart. Phys. {\bf 0906}, 012 (2009)
        [arXiv:0902.2006 [astro-ph.CO]].
\bibitem{Avgoustidis2010} A. Avgoustidis, C. Burrage, J. Redondo, L. Verde and
           R. Jimenez,  J. Cosmol. Astropart. Phys. {\bf 1010}, 024 (2010) [arXiv:1004.2053
          [astro-ph.CO]].
\bibitem{Jaeckel2010} J. Jaeckel and A. Ringwald,  Ann. Rev. Nucl. Part. Sci. {\bf 60}, 405
           (2010) [arXiv:1002.0329 [hep-ph]].
\bibitem{bassett} B. A. Bassett and M. Kunz, \apj {\bf 607}, 661 (2004);
               B. A. Bassett and M. Kunz, Phys. Rev. D {\bf 69}, 101305 (2004).
\bibitem{DeBernardis2006} F. DeBernardis , E. Giusarma and A. Melchiorri,  Int. J. Mod. Phys. D {\bf 15}, 759 (2006).
\bibitem{Reese2002} E. D. Reese, J. E. Carlstrom, M. Joy, J. J. Mohr, L. Grego,
and W. L. Holzapfel, Astrophys. J. {\bf 581}, 53 (2002).
\bibitem{Boname06}       M.  Bonamenteet et al.,  \apj {\bf 647}, 25 (2006).

\bibitem{Lazkoz2008} R. Lazkoz, S. Nesseris and L. Perivolaropoulos,  J. Cosmol. Astropart. Phys. {\bf 07}  012 (2008).



\bibitem{DeFilippis05}   E. De Filippis, M. Sereno, M. W. Bautz and  G. Longo,   \apj {\bf 625}, 108 (2005).
\bibitem{holanda2010}    R. F. L. Holanda, J. A. S. Lima and  M. B. Ribeiro,  Astron. Astrophys. {\bf 528}, L14 (2011).
\bibitem{holanda20103}   R. F. L. Holanda, J. A. S. Lima and  M. B. Ribeiro,  \apj {\bf 722}, 233 (2010).
\bibitem{Hicken2009} M. Hicken et al., Astrophys. J. {\bf 700}, 1097 (2009).
\bibitem{Li2011} Z. Li, P. Wu  and H. Yu,   \apj {\bf 729}, L14 (2011).
\bibitem{Meng2012} X. L. Meng, T. J. Zhang and H. Zhan, \apj {\bf 745}, 98 (2012).
\bibitem{Wu2015}P. Wu, Z. Li, X. Liu  and H. Yu,   Phys. Rev. D {\bf 92}, 023520 (2015).


\bibitem{Ellis2013} G. F. R. Ellis, R. Poltis, J. P.Uzan and A. Weltman,  Phys. Rev. D {\bf 87}, 103530 (2013).

\bibitem{Goncalves2012} R. S. Gon{\c{c}}alves, R. F. L. Holanda and J. S. Alcaniz,  Mon. Not. R. Astron. Soc. {\bf 420}, 43 (2012).
\bibitem{Cao2012}  S. Cao, et al.,  J. Cosmol. Astropart. Phys. {\bf 03}, 016 (2012).
\bibitem{Cao2015} S. Cao et. al., \apj {\bf 806}  185 (2015) [arXiv:1509.07649].
\bibitem{Balmes}I. Balmes and P. S. Corasaniti,  Mon. Not. R. Astron. Soc. {\bf 431}  1528, (2013)[arXiv:1206.5801].
\bibitem{Rana} A. Rana, et al., J. Cosmol. Astropart. Phys.,  {\bf 7}, (7) (2017).
\bibitem{Ruan} C. Ruan, F. Melia and T. Zhang, [arXiv:1808.09331 [astro-ph.CO]].

\bibitem{avtidisgous}    A. Avgoustidis, et al.,  J. Cosmol. Astropart. Phys. {\bf1010}, 024 (2010).
\bibitem{Stern2010}      D. Stern,  R. Jimenez,  M. Kamionkowski and  S. A. Stanford,  J. Cosmol. Astropart. Phys. {\bf 1002}, 008 (2010).
\bibitem{Holanda2012a}  R. F. L. Holanda, R. S. Gon{\c{c}}alves and J. S. Alcaniz, J. Cosmol. Astropart. Phys. {\bf 06}, 022 (2012).
\bibitem{Santos2015} S. Santos-da-Costa, V. C. Busti and R. F. L. Holanda,  J. Cosmol. Astropart. Phys. {\bf 10}, 061 (2015).
\bibitem{Holanda2012} R. F. L. Holanda, J. C. Carvalho and J. S. Alcaniz,  J. Cosmol. Astropart. Phys.  {\bf 1304},  027 (2013). 

\bibitem{Liao2011} K. Liao, Z. Li, J. Ming and Z. Zhu,  Phys. Lett. B {\bf 718},  1166-1170 (2013).

\bibitem{Liao2016} K. Liao, et al.,  \apj {\bf 822}, 74 (2016).
\bibitem{Holanda20171} R. F. L. Holanda, V. C. Busti F. S. Lima and J. S. Alcaniz, J. Cosmol. Astropart. Phys. {\bf 1709}, no. 09, 039 (2017).
\bibitem{fuxiangyun} X. Y. Fu et al.,  Research in Astron. Astrophys.,  {\bf 8}, 895 (2011).

\bibitem{Fu2017} X. Fu and P. Li, Int. J. Mod. Phys. D {\bf 26}, no. 9, 1750097 (2017)
           [arXiv:1702.03626 [gr-qc]].
\bibitem{Liang2013} N. Liang, Z. Li, P. Wu, S. Cao, K. Liao and Z. H. Zhu, Mon.
Not. R. Astron. Soc. {\bf 436}, 1017 (2013).

\bibitem{Tolman1930} R. C. Tolman, Proc. Natl. Acad. Sci. {\bf 16}, 511 (1930).

\bibitem{Abbott} B. P. Abbott et al. [LIGO Scientific and Virgo Collab-
           orations], Phys. Rev. Lett. 119, no. {\bf 16}, 161101 (2017)
            [arXiv:1710.05832 [gr-qc]].
\bibitem{Abbott2} B. P. Abbott et al. [LIGO Scientific and Virgo and Fermi-
           GBM and INTEGRAL Collaborations],  Astrophys. J.
              {\bf 848}, no. 2, L13 (2017) [arXiv:1710.05834 [astro-ph.HE]].
\bibitem{Daz} M. C. Daz, et al. [TOROS Collaboration],   Astrophys. J. Lett. {\bf 848}, no. 2, 29 (2017)
           [arXiv:1710.05844 [astro-ph.HE]].
\bibitem{Cowperthwaite} P. S. Cowperthwaite et al., Astrophys.
            J. {\bf 848}, no. 2, L17 (2017).
\bibitem{Schutz} B. F. Schutz, Nature {\bf 323}, 310 (1986).

\bibitem{Zhao2011} W. Zhao, C. VanDenBroeck, D. Baskaran, and T. G. F. Li, Phys. Rev. D {\bf 83}, 023005 ( 2011).
\bibitem{Nissanke2010}  S. Nissanke,  D. E. Holz, S. A. Hughes,  N. Dalal, and   J. L. Sievers,
\apj {\bf 725}, 496 (2010).
\bibitem{Cai2017} R. G. Cai and T. Yang,  Phys. Rev. D {\bf 95}, 044024 (2017).
\bibitem{Belgacem2018} E. Belgacem, Y. Dirian, S. Foffa and M. Maggiore, Phys. Rev. D 97, 104066 (2018).
\bibitem{Saltas2014} I. D. Saltas, I. Sawicki, L. Amendola, and M. Kunz, Phys. Rev. Lett., {\bf 113}, 191101 (2014).

\bibitem{Pozzo201217} W. Del Pozzo,  Phys. Rev. D {\bf 86}, 043011 (2012);
                 W. Del Pozzo,  T. G. F. Li, and  C. Messenger, Phys. Rev. D {\bf 95}, 043502 (2017).
\bibitem{Cai2016}  R. G. Cai,  Z. K. Guo, and  T. Yang,   Phys. Rev. D {\bf 93}, 043517 (2016).
\bibitem{Liao2017}  K. Liao, X. Fan, X. Ding,  M. Biesiada, and  Z. Zhu,  Nature
                Communications {\bf 8}, 1148 (2017).
\bibitem{Wei2017} J. Wei, and X. Wu,    Mon. Not. R. Astron. Soc. {\bf 472}, 2906 (2017);
                J. Wei, and X. Wu, and H. Gao,    \apj {\bf 860}, L7 (2018).
\bibitem{Wang2018} L. Wang,  X. Zhang, J. Zhang and  X. Zhang,   Phys. Lett. B {\bf 782},  87 (2018).
\bibitem{Wei2018} J. Wei,  ArXiv e-prints, arXiv:1806.09781 (2018).
\bibitem{Cai2018} R. G. Cai, T. B. Liu,  X. W. Liu, S. J. Wang  and  T. Yang,
Phys. Rev. D {\bf 97}, 103005 (2018).
\bibitem{Zhao2018}  W. Zhao,  B. S. Wright and B. Li,  ArXiv e-prints, arXiv:1804.03066 (2018).
\bibitem{Yang2017} T. Yang,  R. F. L. Holanda  and   B. Hu,  ArXiv e-prints, arXiv:1710.10929 (2017).
\bibitem{Bassett2010} B. A. Bassett and R. Hlozek,  edited by P.
Ruiz-Lapuente, (Cambridge University Press, Cambridge,
2010), ISBN-13: 9780521518888.


\bibitem{Suwa2003} T. Suwa, A. Habe,  K. Yoshikawa, and  T. Okamoto,   \apj, {\bf 588}, 7 (2003);  Y. P. Jing, and Y. Suto,  \apj, {\bf 574}, 538 (2002).
\bibitem{Allen2004}   S. W. Allen, R. W. Schmidt,  H. Ebeling, A. C. Fabian, and L. van Speybroeck,
 Mon. Not. R. Astron. Soc., {\bf 353}, 457 (2004).
\bibitem{SZE} R. A. Sunyaev, and  Y. B. Zel¡¯dovich,   Comments on Astrophysics and
Space Physics, {\bf 4}, 173 (1972); J. E. Carlstrom,  G. P. Holder, and E. D. Reese, 2002, ARA\&A, 40, 643.



\bibitem{Reese20023} M. Ribeiro   and W. Stoeger,    \apj {\bf 592}, 1 (2003).

\bibitem{Blake2012} C. Blake, S. Brough, M. Colless et al., Mon. Not. R. Astron. Soc. {\bf 425}, 405 (2012).
\bibitem{Xu2013} X. Xu, A. J. Cuesta, N. Padmanabhan, D. J. Eisenstein, and
C. K. McBride, Mon. Not. R. Astron. Soc. {\bf 431}, 2834 (2013).
\bibitem{Samushia2014} L. Samushia, B. A. Reid, M. White et al., Mon. Not. R.
Astron. Soc. {\bf 439}, 3504 (2014).

\bibitem{Abadie2010a} J. Abadie et al. [LIGO Scientific Collabora-
tion], Nucl. Instrum. Meth. A {\bf 624}, 223 (2010)
doi:10.1016/j.nima.2010.07.089 [arXiv:1007.3973 [gr-
qc]].

\bibitem{Plank2015} P. A. R. Ade et al. [Planck Collaboration], Planck 2015
results. XIII. Cosmological parameters, Astron. Astro-
phys. {\bf 594}, A13 (2016) [arXiv:1502.01589 [astro-ph.CO]].


\bibitem{Bevington2003} P. R. Bevington and  D. K. Robinson,  Data reduction and error analysis
for the physical sciences, 3rd ed., by Philip R. Bevington, and Keith
D. Robinson. Boston, MA: McGraw-Hill, ISBN 0-07-247227-8, (2003).

\end{thebibliography}
\end{document}